\documentclass[prl,twocolumn,superscriptaddress]{revtex4}
\usepackage{graphicx}
\begin{document}
\newcommand{\be}{\begin{equation}}
\newcommand{\ee}{\end{equation}}
\newcommand{\bea}{\begin{eqnarray}}
\newcommand{\eea}{\end{eqnarray}}
\title{Brane Vector Phenomenology}
\author{T.E. Clark}
\email{clark@physics.purdue.edu}
\affiliation{Department of Physics,
Purdue University,
West Lafayette, IN 47907-2036, U.S.A.}
\author{S.T. Love}
\email{love@physics.purdue.edu}
\affiliation{Department of Physics,
Purdue University,
West Lafayette, IN 47907-2036, U.S.A.}
\author{Muneto Nitta}
\email{ nitta@phys-h.keio.ac.jp}
\affiliation{Department of Physics,
Keio University,
Hiyoshi, Yokohoma, Kanagawa, 223-8521, Japan}
\author{T. ter Veldhuis}
\email{terveldhuis@macalester.edu}
\affiliation{Department of Physics \& Astronomy,
Macalester College,
Saint Paul, MN 55105-1899, U.S.A.}
\author{C. Xiong}
\email{xiong@purdue.edu}
\affiliation{Department of Physics,
Purdue University,
West Lafayette, IN 47907-2036, U.S.A.}
\begin{abstract}
Local oscillations of the brane world are manifested as masssive vector  fields. Their coupling to the Standard Model can be obtained using the method of nonlinear realizations of the spontaneously broken higher dimensional space-time symmetries, and to an extent, are model independent.  Phenomenological limits on these vector field parameters are obtained using LEP collider data and dark matter constraints.
\end{abstract}

\maketitle

The quanta associated with the vibrations of a flexible brane world can signal the existence of extra space dimensions.  One type of excitation common to all such brane world models is a massive vector\cite{local}. These massive vectors can be produced in colliders and as such their properties can be constrained. Moreover, if they are stable, they are  dark matter candidates and consequently subject to the limits set by the various corresponding searches. In this letter, we examine the phenomenological restrictions on these  properties in a model independent manner.  

The brane world spontaneously breaks the higher dimensional space-time symmetries down to those of the world volume and its covolume.  The method of nonlinear realizations of spontaneously broken symmetries has been used to construct the Nambu-Goto action for the consequent Nambu-Goldstone bosons of this spontaneous space-time symmetry breakdown\cite{coset}.  These scalar fields describe the oscillations of the brane into the covolume.  When the back reaction of the geometry is included with the gravitational fields being dynamic, the gravitational Higgs mechanism occurs with the brane oscillation (branon\cite{branon}) Nambu-Goldstone boson scalar fields being eaten by the dynamic zero mode higher dimensional vector fields.  This results in brane oscillation world volume massive vector (Proca) fields appearing in the brane world, denoted $X_\mu^i$, with $i=1,2,\ldots, N$, where $N$ is the dimension of the covolume, and the Greek indices are those of the world volume, $\mu =0,1,2,3$.  

The coset method has been extended to include the treatment of spontaneously broken local space-time symmetries.  This has given the higher dimensional invariant coupling of the massive brane vector fields to themselves, the graviton, and the Standard Model fields. It is the latter which we consider in  this letter and hence we focus on the gauging of the spatial translations in the extra dimensions.  Expanding the action in terms of the number of brane vectors, a $D=(4+N)$-dimensional invariant world volume effective action is given by
\bea
{\cal L}_{\rm effective}&=& {\cal L}_{\rm SM} -\frac{1}{4}F_{\mu\nu}^i F_i^{\mu\nu} +\frac{1}{2}M_X^2 X_\mu^i X_i^\mu \cr
 & &  +\frac{1}{2}\frac{M_X^2}{F_X^4}X^\mu_i T^{\rm SM}_{\mu\nu} X^\nu_i \cr 
 & &  +\frac{M_X^2}{F_X^4}\left(K_1 B_{\mu\nu} +K_2 \tilde{B}_{\mu\nu} \right) \partial^{\mu} X^{\rho}_i \partial_{\rho} X^{\nu}_i ~. 
\label{effaction}
\eea
Here $F_X^4$ is the brane tension, $M_X$ is the vector mass and $K_1, K_2$ are dimensionless parameters of the effective Lagrangian. The coupling to $T^{\rm SM}_{\mu\nu}$, the Standard Model energy-momentum tensor, has its origin in the induced metric on the brane which accounts for the brane motion into the extra dimensions. It is the action one obtains in the minimal case of gauging the globally invariant lowest order coupling of the branon Nambu-Goldstone bosons to the Standard Model fields in which case only the Standard Model energy-momentum tensor, $T^{\rm SM}_{\mu\nu}$, coupling term is present.  From the general effective action point of view, however, the extrinsic curvature allows additional new interactions with the Standard Model fields.  Since the coupling must be $SU(3)\times SU(2)\times U(1)$ invariant, so must the Standard Model operators.  The lowest dimension such terms are the hypercharge field strength tensor, $B_{\mu\nu}$, and its dual, $\tilde{B}_{\mu\nu}$. A coupling to the invariant Higgs bilinear $H^\dagger H$ is also possible and leads to possible new invisible Higgs decay modes which will be discussed elsewhere. 

Note that the above effective Lagrangian exhibits an unboken global $SO(N)$ symmetry which reflects the various equivalent ways of embedding the brane in the higher dimensional space. Since the brane vector fields carry the unbroken global $SO(N)$ charge, they must appear pairwise in any  $SO(N)$ invariant effective action. Consequently,  in such an isotropic codimension brane world model, the brane vector particles are stable.

\begin{figure*}
\hspace{-3.5 cm}
\begin{minipage}{5 cm}
\includegraphics[width=3.25in]{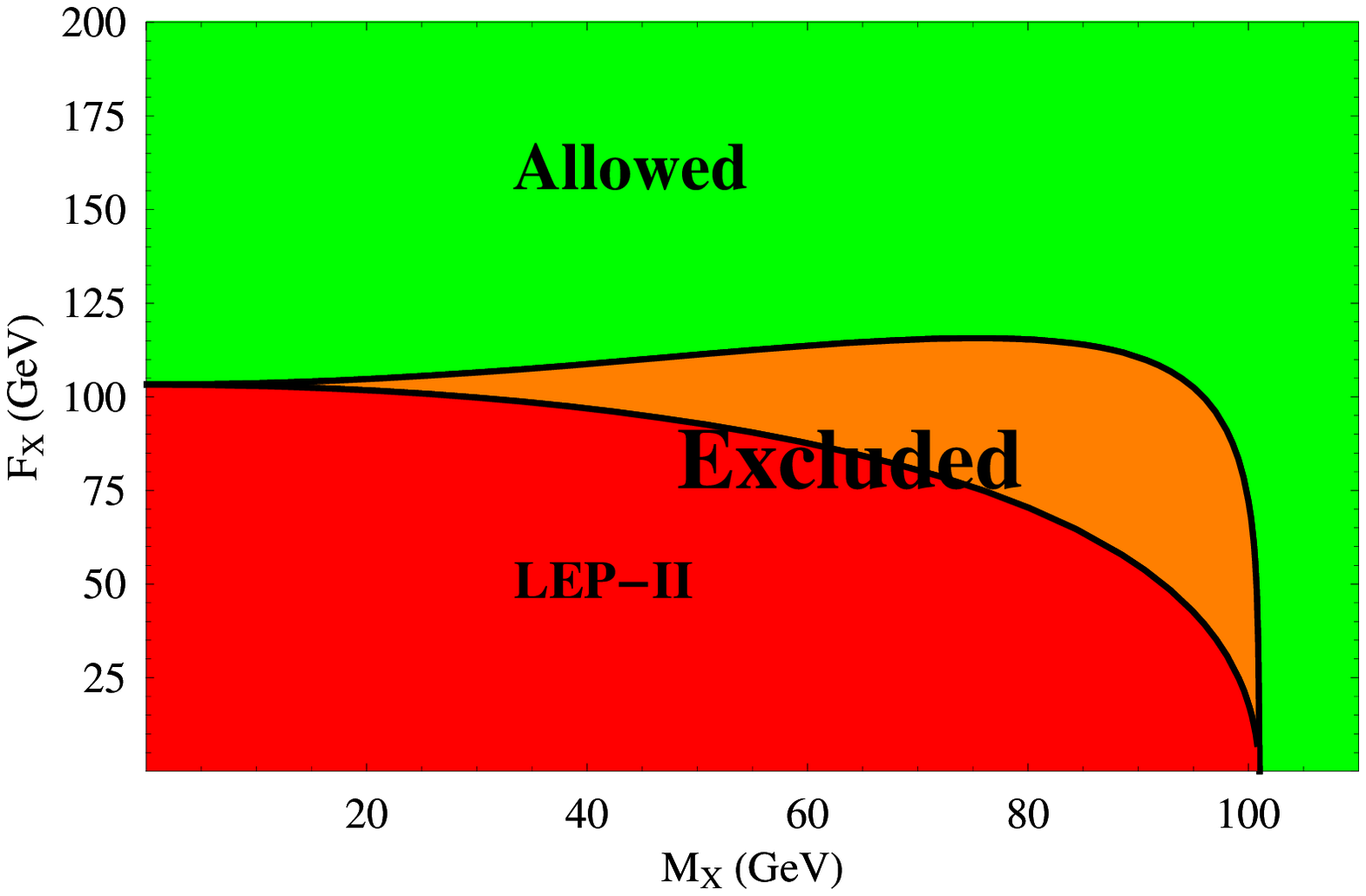}
\end{minipage}
\hspace{4 cm}
\begin{minipage}{5 cm}
\includegraphics[width=3.25in]{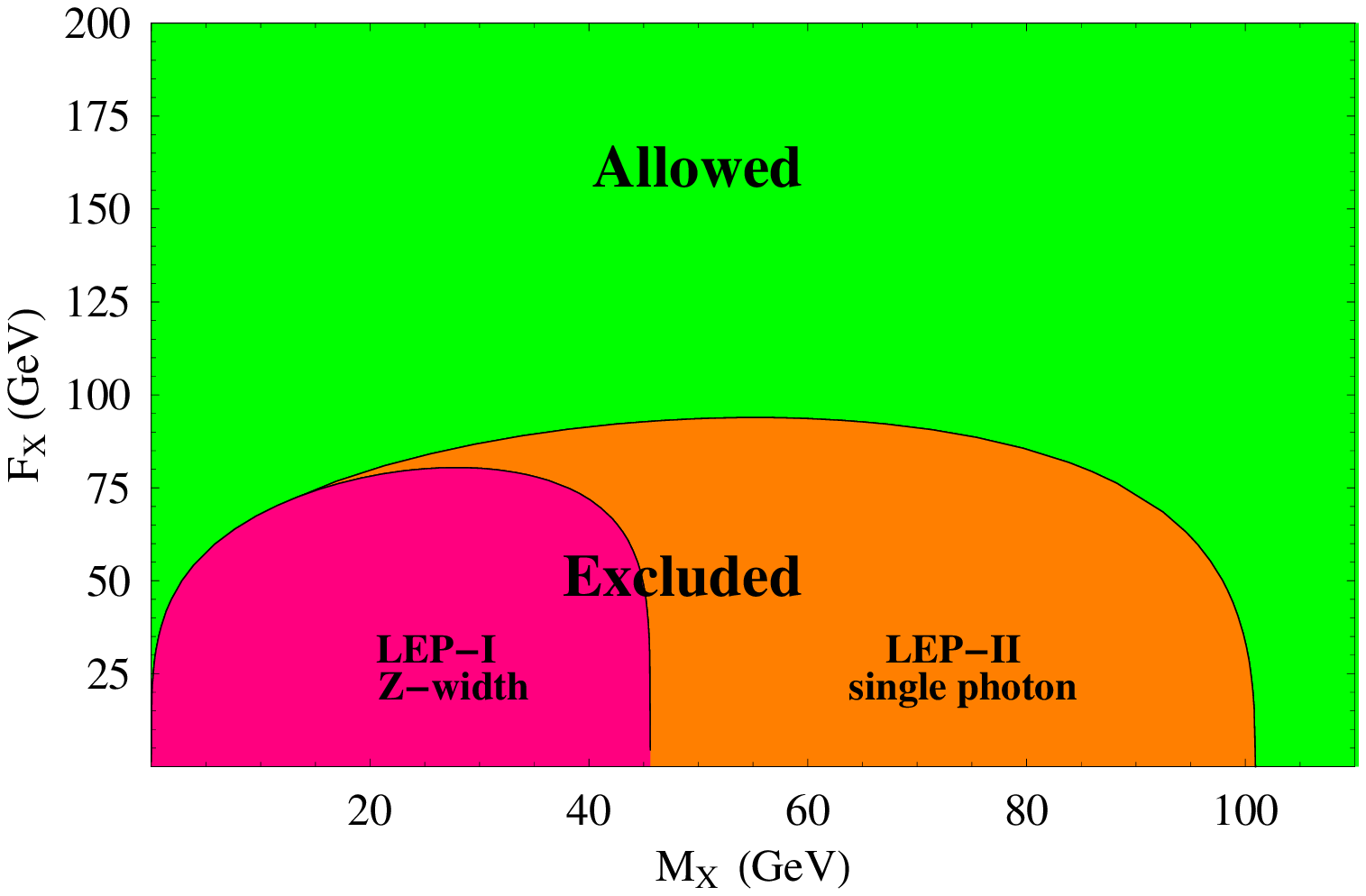}
\end{minipage}
\caption{LEP excluded regions of brane vector parameter space $M_X$-$F_X$.  The left plot is the case where the brane vectors couple to the Standard Model energy-momentum tensor with no extrisic curvature coupling $K_1 =0=K_2$.  The right plot is when the extrinsic curvature coupling is dominant and the coupling to the energy-momentum tensor is neglected.  In this case the coupling is just to the weak hypercharge field strength tensor with strength $K_1 =K_2=1$.}
\end{figure*}

Data from the lepton colliders LEP-I\cite{LEP1} and LEP-II\cite{LEP2} can be employed to give excluded and allowed regions for the $M_X$-$F_X$ (as well as codimension $N$) parameter space.  In particular the annihilation of a fermion and anti-fermion to produce a single photon and two brane vectors which escape the detector as missing energy is used to delineate the regions of this parameter space. In addition, the limit on the allowed width of invisible $Z$-decays, $\Gamma_{Z\rightarrow XX}\leq 2$ MeV, provides additional constraints on the parameter space and in particular the extrinsic curvature coupling constants.  Data from LEP-II for $e^+ e^- \rightarrow \gamma + XX \rightarrow \gamma +\rlap{E}{/}$~ is in agreement with the expected results from the Standard Model.  This lack of discovery places a bound on the contribution of the brane vectors to the missing energy cross-section.  The number of brane vector missing energy events must be less than 5 sigma of the Standard Model background $\sigma_{\gamma XX} {\cal L}_{\rm LEP-II} \leq 5 \sqrt{N_{\rm SM bkgrnd}}.$  Using the integrated luminosity for LEP-II, ${\cal L}=138.8~{\rm pb}^{-1}$, at an average center of mass energy of $\sqrt{s}=206$ GeV, a bound on the brane vector parameters is obtained as $\sigma_{\gamma XX}\leq 0.45$ pb. In order to distinguish the constraints that the various interaction terms place on the brane tension and brane vector mass the plot on the left of Fig. 1 includes only the interaction with the Standard Model energy-momentum tensor with the extrinsic curvature derivatively coupled brane vector terms are set to zero, $K_1 =K_2=0$.  The plots are quite insensitive to the value of $N$ for codimension $N=1$ as the line of exclusion varies according to $N^{1/8}$. As such they are presented for $N=1$. The corresponding delineation of parameter space in the global symmetry case containing only the branon scalar (longitudinal vector) contribution is shown as the red colored excluded area with the orange and red region excluded in the case of the brane vector.  
On the other hand, in the plot on the right hand side of Fig. 1 the energy-momentum tensor interaction terms are ignored in comparison to the interactions that are due to the extrinsic curvature with $K_1 = K_2 =1$. In addition the constraints on parameter space due to the LEP-I limits on the invisible $Z$-decay into two brane vectors are depicted in Fig. 1.

\begin{figure}
\begin{center}
\includegraphics[width=3.25in]{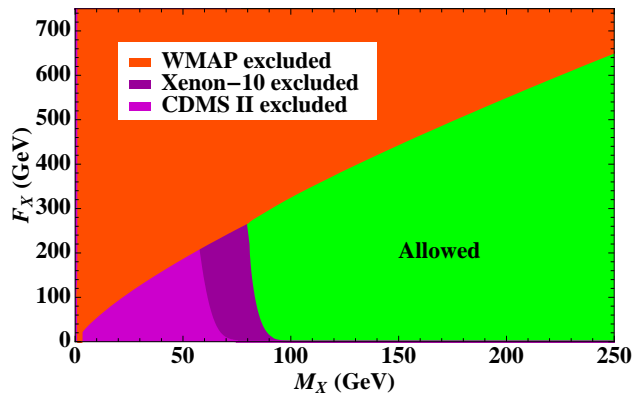}
\end{center}
\caption{Constraints on the effective theory $M_X$-$F_X$ parameter space for a single flavor of brane vectors due to the WMAP dark matter abundance result and the CDMS II and XENON-10 direct dark matter detection data. \label{fig3} }
\end{figure}
Since for this isotropic codimension case, the brane vectors they are stable particles, they are candidates for dark matter.  In order to determine the viability of their candidacy the minimal case of no extrinsic curvature coupling, $K_1=K_2=0$, is considered.  The dark matter constraints on the remaining $M_X$-$F_X$ parameter space are obtained from a combination of the dark mater density result as  reported by the WMAP collaboration \cite{Spergel:2006hy} from their fit of the cosmological parameters to the cosmic microwave anisotropy data,  and the limits on the spin independent  cross-section of dark matter particles scattering off a nucleon as reported by the CDMS II \cite{Akerib:2005kh} and XENON-10 \cite{Angle:2007uj} collaborations from the non-observation of a signal in their direct dark matter detection experiments.
The relic abundance of the brane vector as a function of its mass $M_X$ and the scale $F_X$  is calculated according to standard methods. It is assumed that at some point during the evolution of the Universe the population of brane vectors  is in thermal equilibrium with the population of Standard Model particles. The density of brane vectors  as a function of time then follows from simple thermodynamics as long as they remain in thermal equilibrium. However, at some point in time the expansion rate of the Universe exceeds the annihilation rate of the brane vectors. The brane vector then falls out of thermal equilibrium, and from that moment on its density only changes due to the expansion of the Universe while annihilation effectively ceases. Typically, this freeze out occurs at temperatures that are about $1/20$ times the mass $M_X$. The brane vectors are therefore non-relativistic at freeze out, and consequently contribute to the cold dark matter density. In the numerical calculations, the freeze out temperature is determined by taking into account the non-relativistic annihilation cross-sections of  pairs of brane vectors into all relevant pairs of Standard Model particles, and the corresponding relic brane vector abundance is obtained for each point in parameter space.

Fig. \ref{fig3} shows the dark matter constraints on the $M_X$-$F_X$ parameter space in the case of a single flavor of brane vectors. In the red area of the graph the brane vector annihilation cross-section is so small that the brane vector freezes out too early in the evolution of the Universe and as a consequence its relic density exceeds the WMAP result. In the green area the brane vector relic density is below the WMAP dark matter density result, while at the boundary between the red and green areas the brane vector relic abundance matches the WMAP dark matter density result. In order for the green area of the parameter space to be consistent with the WMAP data, there must exist additional types of dark matter particles beyond the brane vector that are not included in the effective theory in order to make up for the deficit of the brane vector relic abundance as compared to the WMAP dark matter density result. When applying the dark matter direct detection constraints to the parameter space it is  assumed that such additional types of dark matter have vanishing interactions with the nucleon. Under this conservative assumption, the lighter purple area in the graph is excluded by the CDMS II data, while the darker purple area is excluded by the XENON-10 data.

All of the above discussion assumed that all codimensions were isotropic. For any anisotropic codimension, there can be couplings to the Standard Model linear  in $X^\mu$.  Thus, in this case, the $X$ vector acts like a $Z^\prime$ boson and is subject to its various experimental limits\cite{LEP1}. 

\vspace*{1.0in}

The work of TEC, STL and CX was supported in part by the U.S. Department of Energy under grant DE-FG02-91ER40681 (Task B).  The work of TtV was supported in part by a Cottrell Award from the Research Corporation.  TtV would like to thank the theoretical physics group at Purdue University for their hospitality during his sabbatical leave from Macalester College.

\newpage

\begin{thebibliography}{100}
\bibitem{local}
T.E. Clark, S. T. Love, M. Nitta, T. ter Veldhuis and C. Xiong,
{Phys. Rev.} D{\bf 75}, 065028 (2007)  ,[arXiv:hep-th/0612147]; S.T. Love,  {J. Phys.} A: { Math. Theor.} {\bf 40}, 7049  (2007), [arXiv:hep-th/0611199]; T.E. Clark, S.T. Love, M. Nitta and T. ter Veldhuis, { Phys. Rev.} D{\bf 72}, 085014 (2005), [arXiv:hep-th/0506094]. 


\bibitem{coset}T.E. Clark, S.T. Love, M. Nitta, T. ter Veldhuis and C. Xiong, arXiv:hep-th/0703179; T.E. Clark and S.T. Love, { Phys. Rev.} D{\bf 73}, 025001 (2006), [arXiv:hep-th/0510274]; S.T. Love,  { Mod. Phys. Lett.} A{\bf 20},  2903-2911 (2005), [arXiv/hep-th/0510187]
; T.E. Clark, S.T. Love, M. Nitta and T. ter Veldhuis, { J. Math. Phys.} {\bf 46}, 102304 (2005), [arXiv:hep-th/0501241]
; J. Gomis, K. Kamimura and P. West, { Class. Quant. Grav.}  {\bf 23}, 7369 (2006)[arXiv:hep-th/0607057];
{\it JHEP} {\bf 0610}, 015 (2006); [arXiv:hep-th/0607104]; 
E. A. Ivanov and V. I. Ogievetsky, { Teor. Mat. Fiz.}  {\bf 25}, 164 (1975).




\bibitem{branon}
Creminelli and Strumia, { Nucl. Phys.} B{\bf 596}, 125 (2001); 
Alcaraz, Cembranos, Dobabo and Maroto, { Phys. Rev.} D{\bf 67}, 075010 (2003);  
S. Mele, EPS-HEP05, 153. 


\bibitem {LEP1}W.-M. Yao et al., Particle Data Group,  { J. Phys. } G{\bf 33}, (2006). 

\bibitem {LEP2}P. Achard et al., L3 Collaboration,  { Phys. Lett.} B{\bf 597}, 145 (2004).

\bibitem {Spergel:2006hy}D.N. Spergel et al. [WMAP Collaboration], Astrophys. J. Suppl. {\bf 170}, 370 (2007); WMAP web sites, http://lambda.gsfc.nasa.gov/product/map/.

\bibitem {Akerib:2005kh}D.S. Akerib et al., [CDMS Collaboration], { Phys. Rev. Lett.}  {\bf 96}, 011302 (2006).

\bibitem {Angle:2007uj}J. Angle et al., [XENON Collaboration], arXiv:astro-ph/07060039.


\end{thebibliography}
\end{document}